\documentclass{article}
\usepackage{jheppub,mymacros}
\title{{C=Anything and the switchback effect in\\ Schwarzschild-de Sitter space}}
\author[a]{Sergio E. Aguilar-Gutierrez}
\affiliation[a]{Institute for Theoretical Physics, KU Leuven, Celestijnenlaan 200D, B-3001 Leuven, Belgium}
\emailAdd{sergio.ernesto.aguilar@gmail.com}
\abstract{We investigate observables within the framework of the codimension-one C=Anything (CAny) proposal for Schwarzschild-de Sitter (SdS) space under the influence of shockwave sources. Within the proposal, there is a set of time-reversal invariant observables that display the same rate of growth at early and late times for a background with or without shockwave sources. Once we introduce shockwaves in the weak gravitational coupling regime, there is a decrease in the late-time complexity growth due to cancellations with early-time perturbations, known as the switchback effect. The result shows that some CAny observables in SdS may reproduce the same type of behavior found in anti-de Sitter black holes. We comment on how our results might guide us to new explorations in the putative quantum mechanical theory.}
\begin{document}

\maketitle
\section{Introduction}

Recently, a lot of interest in quantum information theoretic notions has surfaced in an effort to characterize semiclassical gravitational observables. Particularly, holographic complexity has been used to study bulk gravitational dynamics that can probe regions inaccessible to entanglement entropy \cite{Susskind:2014moa,Susskind:2014rva,Stanford:2014jda}, and whose variation can also reproduce gravitational equations of motion  \cite{Czech:2017ryf,Caputa:2018kdj,Susskind:2019ddc,Pedraza:2021mkh,Pedraza:2021fgp,Pedraza:2022dqi}. This notion is also expected to reproduce the properties for the computational complexity of a quantum circuit model\footnote{See \cite{Chapman:2021jbh} for a recent review.} of the conformal field theory (CFT) dual to asymptotically Anti-de Sitter (AdS) spacetimes \cite{Susskind:2014rva}. This translates to robust features captured by a large family of proposals, starting with the Complexity=Volume (CV) \cite{Stanford:2014jda}, Complexity=Action (CA) \cite{Brown:2015bva, Brown:2015lvg}, Complexity=Spacetime Volume (CV2.0) \cite{Couch:2016exn}, and a recent generalization known as the Complexity=Anything (CAny) proposal \cite{Belin:2021bga,Belin:2022xmt}. There are two defining features for the CAny observables in AdS black holes: (i) a late boundary time linear growth, and (ii) the switchback effect. The latter is a characteristic decrease in the late-time linear growth once perturbations in the geometry are introduced due to energy pulses.

Perhaps, one of the most exciting aspects about this family of spacetime probes is to characterize the properties of cosmological backgrounds, and in particular for de Sitter (dS) space. The nature of the microscopic degrees of freedom encoded inside the cosmological horizon remains mysterious (see \cite{Galante:2023uyf} for a recent review).

A recent proposal in dS holography, known as stretched horizon holography \cite{Susskind:2021esx,Shaghoulian:2022fop,Susskind:2022dfz,Lin:2022rbf,Lin:2022nss,Bhattacharjee:2022ave,Rahman:2022jsf,Susskind:2023hnj,Susskind:2023rxm,Franken:2023pni,Lin:2023trc}, has sparked many developments in dS space complexity. The stretched horizon is a region in the static patch of dS space where the dual theory is conjectured to be located \cite{Susskind:2021esx}. One may therefore perform gravitational dressings with respect to the stretched horizon to probe the dual degrees of freedom. Although the precise location is not explicit in this approach, it is expected to be close to the cosmological horizon. This approach has been explored to study holographic complexity proposals in asymptotically dS space \cite{Jorstad:2022mls,Auzzi:2023qbm,Anegawa:2023dad,Baiguera:2023tpt,Anegawa:2023wrk,Aguilar-Gutierrez:2023zqm,Aguilar-Gutierrez:2023tic}. One of the striking features originally found in \cite{Jorstad:2022mls} for the CV, CA, and CV2.0 proposals was that the rate of growth of holographic complexity might diverge a finite times relative to the stretched horizon, denoted by hyperfast growth. The result was associated with the conjecture where the double-scaled Sachdev–Ye–Kitaev (SYK) model is identified as the quantum mechanical dual to dS$_2$ space \cite{Susskind:2021esx}. One may introduce a cutoff surface to perform an analytic continuation allowing for late-time evolution \cite{Jorstad:2022mls}.

However, it was recently found that hyperfast growth is not a universal phenomenon in the space of holographic complexity proposals \cite{Aguilar-Gutierrez:2023zqm}. Instead, a set of the CAny observables may evolve to arbitrarily late (or early) static patch times without introducing regulator surfaces. Notice, however, that it is still unclear what kind of interpretation the different types of CAny observables might have for dS space. To properly define holographic complexity proposals one needs a good understanding of the basic properties that complexity for the dual field theory side should satisfy. Such understanding of the microscopics associated with dS space is still lacking. It is important to learn about the different signals one should look for in a candidate for the dS space holographic model. 

Alternatively, one might find clearer interpretations of holographic complexity when dS space is embedded in a higher dimensional bulk AdS spacetime. This perspective was recently approached by \cite{Aguilar-Gutierrez:2023tic}. They considered a particular type of CAny proposals in a braneworld model consisting of a higher dimensional AdS bulk geometry capped off by a pair of dS space end-of-the-world branes. In this setting, the resulting complexity is associated with the field theory dual living on a brane near the asymptotic boundary of the AdS space. In this case, the CAny proposals with the expected late time growth in the AdS bulk are those that obey the late time growth in the dS braneworld.

On the other hand, one of the defining properties of holographic complexity proposals in the AdS context, the switchback effect, has been recently explored in asymptotically dS spacetime \cite{Anegawa:2023wrk,Baiguera:2023tpt} for the CV, CA, CV2.0 proposals. In this case, there are energy pulses associated with perturbations in the evolution of the putative dual theory residing in the stretched horizon. This can be an important diagnostic if the CAny observables can indeed be associated with complexity consistent with Nielsen's geometric approach in the dual theory \cite{Aguilar-Gutierrez:2023zqm}. However, the switchback effect in the class of CAny observables where the late-time growth in dS space is allowed has not been studied. This is the main goal of our work, to learn new lessons for stretched horizon holography.

To study the switchback effect, we describe the shockwave geometry in asymptotically dS spacetimes. We specialize in Schwazschild-de Sitter (SdS) space, which describes spherically symmetric vacuum solutions to Einstein equations with a positive cosmological constant. We work in the perturbative weak gravity regime to treat the shockwave geometry based on previous findings in \cite{Shenker:2013pqa}, and recently discussed in the context of SdS black holes by \cite{Aguilar-Gutierrez:2023ymx}. As for the observables, we will work with codimension-one CAny proposals evaluated in constant mean curvature (CMC) slices, originally introduced in \cite{Belin:2022xmt}. The mean curvature is one of the main factors distinguishing the proposals that display late time growth from those with hyperfast growth \cite{Aguilar-Gutierrez:2023zqm}.

Our work is focused on alternating early and late-time perturbations in the geometry and the resulting growth of the CAny observables. Our findings indicate that the set of proposals with late-time growth will display the switchback effect. The rate of growth will be determined by the definition of the CAny proposal. In general, however, the manifestation of the switchback effect in the CAny proposals only occurs once we perform a time-reversal symmetric extension, originally hinted in \cite{Aguilar-Gutierrez:2021bns}. We select CMC slices that minimize the CAny proposal and find that the late-time and early-time contributions to complexity growth will partially cancel out. Under these conditions, the analysis of the switchback effect shows great similarity with respect to that of AdS black hole backgrounds. 

Interestingly, the late time growth during the switchback phase is unaffected by the particular location of the stretched horizon.

The structure of the manuscript goes as follows. In Sec. \ref{sec:CAny SdS} we review generalities about the shockwave geometries in SdS spacetimes, as well as the CAny proposals that display the early and late-time growth in SdS spacetime. We introduce some new results regarding the generalizing the set of proposals that reproduce the late time growth of complexity in SdS space of arbitrary mass (below the extremal one). In Sec. \ref{sec:switchback}, we present the results on the switchback effect in SdS spacetime by performing a series of alternating shockwave insertions in the background geometry in the weak gravity regime. Finally, Sec. \ref{sec:conclusions} includes a summary of our findings in this setting and some interesting directions for future research. For the convenience of the reader, we provide an App. \ref{App:details} containing some of the details about the evaluation of the late-time growth of the CAny proposals.

\section{C=Anything in SdS spacetimes}\label{sec:CAny SdS}
In this section, we briefly review basic notions about SdS$_{d+1}$ black holes, the modification in the geometry due to shockwave insertions, and the CAny proposals that we investigate in this work. New results include extending the set of CAny proposals studied in \cite{Aguilar-Gutierrez:2023zqm} for SdS space, to more general observables than volumes of CMC slices. This allows for different types of late-time growth behaviors in $d=2$, $d=3$. We also show that late-time growth persists for arbitrary mass black holes.

\subsection{SdS spacetimes}\label{sec: SdS}
The configuration of interest is SdS$_{d+1}$ space, described by the line element
 \begin{align}
  \rmd s^2 &= -f(r) \rmd t_{L/R}^2 + \frac{\rmd r^2}{f(r)} + r^2 \rmd \Omega_{d-1}^2 \, , \label{metric1} \\ 
  f (r) &= 1 - \frac{r^2}{\ell^2} - \frac{2\mu}{r^{d-2}}~,\quad \mu\equiv \frac{16 \pi G_N M  }{(d-1) \Omega_{d-1} r^{d-2}}~, \label{eq:blackeningfactor}
  \end{align}
with $\ell^2=d(d-1) /(2 \Lambda)$;\footnote{Through the rest of the work, we use rescaled coordinates where $\ell=1$.} $\Lambda>0$ is the cosmological constant; $M$ parametrizes the mass of the black hole; $G_N$ is Newton's constant; $\Omega_{d-1}=2\pi^{d/2}/ \Gamma(d/2)$ is the volume of a unit ($d-1$)-sphere; $\mu\in [0,\,\mu_N]$. The case where $\mu=\mu_N$, with
\begin{equation}
    \mu_N=\frac{1}{d}\qty(\frac{d-2}{d})^{\frac{d-2}{2}}~,
\end{equation}
describes the most massive black hole supported in dS space, which we refer to as the extremal SdS limit. Meanwhile, $\mu=0$, reproduces dS$_{d+1}$ space.

We will be interested in describing shockwaves in the geometry. It's convenient to use Kruskal coordinates, defined by
\begin{equation}\label{eq:Kruskal}
    \begin{aligned}
    U_{\rm b, \rm c}&=\rme^{\frac{f'(r_{\rm b, \rm c})}{2}\left(r_{*}(r)-t\right)}~,\\
    V_{\rm b, \rm c}&=-\rme^{\frac{f'(r_{\rm b, \rm c})}{2}\left(r_{*}(r)+t\right)}\,.
\end{aligned}
\end{equation}
where the subindices ``b, c" denote coordinates on the black hole and inflating patches. These patches are centered at the horizons $r_{\rm b, \rm c}$, and cover the range $0\leq r< r_{\mathcal{O}}$ and $r_{\mathcal{O}} \leq r < \infty$ respectively, with $r_{\mathcal{O}}$ a reference point, which we take as the location of the static patch observer. In the present work, we will focus on the inflating region, as we aim to probe it with CAny observables. We replace $U_c$, $V_c\rightarrow$ $U$, $V$ in what follows. We can then express (\ref{metric1}) as
\begin{equation}
    \rmd s^2=-\frac{4 f(r)}{f'(r_{\rm c})}\rme^{-f'(r_{\rm c})r_{*}(r)}\rmd U\,\rmd V +r^2 \rmd \Omega_{d-1}^2
\end{equation}
where $r_{\mathcal{O}} \leq r < \infty$, and $r_{*}=\int \frac{\rmd r}{f(r)}$ is the tortoise coordinate.

\begin{figure}[t!]
    \centering
\begin{minipage}{.5\textwidth}
        \centering
        \includegraphics[height=0.5\textwidth]{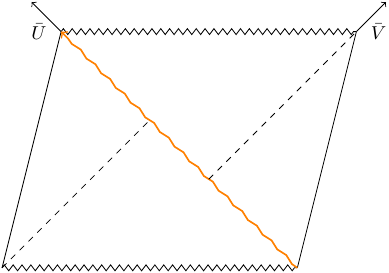}\vspace{0.5cm}\\
        \includegraphics[height=0.5\textwidth]{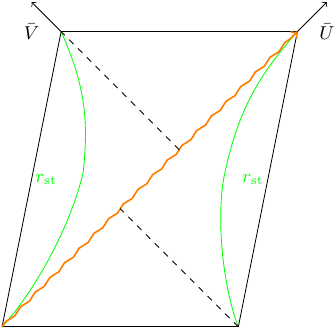}\vspace{0.5cm}\\
        \includegraphics[height=0.5\textwidth]{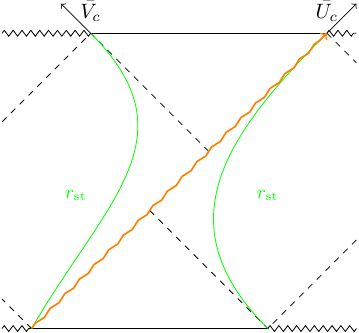}
    \end{minipage}\begin{minipage}{0.5\textwidth}
        \centering
        \includegraphics[height=0.5\textwidth]{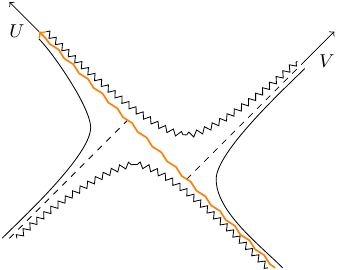}\vspace{0.5cm}\\
        \includegraphics[height=0.5\textwidth]{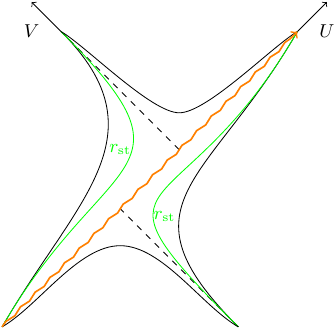}\vspace{0.5cm}\\
        \includegraphics[height=0.5\textwidth]{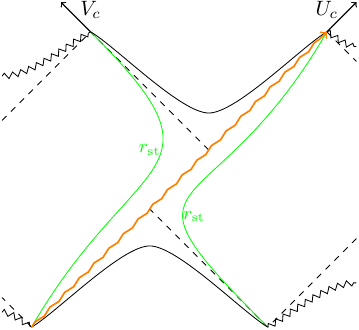}
    \end{minipage}
    \caption{Shockwave geometry for an insertation along $U=0$ (orange wavy line) in: Schwarzchild-AdS$_{d+1\geq4}$ (SAdS$_{d+1\geq4}$) space (\textit{top}, based on \cite{Shenker:2013pqa}), SdS$_3$ space (\textit{middle}); and SdS$_{d+1\geq4}$ space (\textit{bottom}). \textit{Left column}: Penrose diagrams in the $\bar{U}$, $\bar{V}$ coordinates defined in (\ref{eq:conformal coord}). \textit{Right column}: Diagrams in Kruskal coordinates (\ref{eq:metric SW alpha}) with the modification $\alpha\rightarrow-\alpha$ in SAdS$_{d+1\geq4}$ case. In the SdS geometries, the stretched horizon (in green) is shown at a fixed location $r=r_{\rm st}$. The cosmological and black hole horizons are shown with the dashed lines. In all cases, the $U$, $V$-axis are displayed with black arrows.}
    \label{fig:SW SdS}
    \end{figure}

Once we add shockwave perturbations, the geometry becomes distorted, see Fig. \ref{fig:SW SdS}. Notice that as a result of the shockwave in the inflating region of SdS space, the previously causally disconnected static patches become causally connected \cite{Gao:2000ga}. Moreover, the location of the stretched horizon can be modified due to the shift displayed in the diagrams. Although stretched horizon holography fixes the location of the dual theory to be located at a constant $r$ surface in the static patch, it remains unknown how the theory should behave under shockwave perturbations. We will be considering the case where the shockwaves are sent through $U=0$ and the stretched horizon remains fixed at a constant $r=r_{\rm st}$ coordinate\footnote{Similar considerations have been carried out in \cite{Baiguera:2023tpt,Anegawa:2023dad}, as well as alternative proposals.}, for which time evolution along the stretched horizon is continuous.
%In the different diagrams, we employ: (\textit{left column}) $\tilde{U}$, $\tilde{V}$ coordinates in (\ref{eq:conformal coord}), and (\textit{right column}) Kruskal coordinates (\ref{eq:Kruskal}).
We are mainly interested in the metric under shockwave perturbations in the weak gravitational coupling regime to define the notions of energy below. We consider an SdS black hole with mass $M$ absorbs a shell of matter with mass $E\ll M$ along the surface 
\begin{equation}
    U=U_0=\rme^{\frac{f'(r_{\rm c})}{2}\left(r_{*}(r_{\mathcal{O}})-t_0\right)}~,
\end{equation}
with $t_0$ the static time shockwave insertion with respect to $r_{\mathcal{O}}$.

The SdS black hole after the shockwave has mass $M - E$ in (\ref{metric1}) for matter obeying the NEC \cite{Baiguera:2023tpt,Aguilar-Gutierrez:2023ymx}. We glue the coordinates along a shell $U,\, V$ to the past of the shell with those to the future, denoted by $\tilde{U},\, \tilde{V}$. The resulting cosmological line element for SdS black holes \cite{Aguilar-Gutierrez:2023ymx,Shenker:2013pqa}:
\begin{equation}\label{eq:tilde metric}
    \rmd s^2=-\frac{4 \tilde{f}(r)}{\tilde{f}'(r_{\rm c})}\rme^{-\tilde{f}'(\tilde{r}_{\rm c})\tilde{r}_{*}(r)}\rmd \tilde{U}\rmd \tilde{V} +r^2 \rmd \Omega_{d-1}^2
\end{equation}
where tilded quantities are given by the replacement of $M\rightarrow M-E$ in the untilded ones. In the inflating patch, the shift along the $V$ coordinate can be described by a shift in the coordinate
\begin{equation}
    \tilde{V}=V-\alpha~.
\end{equation}
The NEC also imposes that $\alpha\geq0$ \cite{Aalsma:2021kle}; while $E\ll M$ guarantees we work in the $\alpha\ll1$ limit\footnote{See \cite{Aguilar-Gutierrez:2023ymx} for remarks on the approximation for the extremal SdS limit.}. Importantly, this allows for the static patches in Fig. \ref{fig:SW SdS} to become causally connected, as it has been shown rigorously by Gao and Wald \cite{Gao:2000ga}, in contrast to crunching geometries where the shift in $\alpha$ takes the opposite sign for matter satisfying the NEC.

We will express the shift parameter as
\begin{equation}\label{eq:scrambling cosmo}
\begin{aligned}
    \alpha=2\rme^{-\frac{f'(r_c)}{2}(t_c^{(*)}\pm t_0)}~.
    \end{aligned}
\end{equation}
Here the $\pm$ sign depends on whether the shockwave is left or right moving, and the cosmological scrambling time $t_c^{(*)}$ will be defined through this relation.%\cite{Stanford:2014jda}

We will take the shockwave close to the cosmological horizon and set $U_0=0$ in the following. (\ref{eq:tilde metric}) transforms into
\begin{align}\label{eq:metric SW alpha}
        &\rmd s^2=-2A\qty(U[V-\alpha\Theta(U)])
    \rmd U\rmd V+B\qty(U[V-\alpha\Theta(U)])\rmd \Omega_{d-1}^2~,\\
    &A\qty(UV)=-\frac{2}{UV}\frac{f(r)}{f'(r_c)^2}~,\quad B\qty(UV)=r^2~.\label{eq:factors a b}
\end{align}
Moreover, one can consider the change of coordinates
\begin{equation}\label{eq:conformal coord}
    \bar{U}=\arctan U,\quad \bar{V}=\arctan V~,
\end{equation}
for the respective Penrose diagram (see Fig \ref{fig:SW SdS}).

\subsection{C=Anything: CMC slices}\label{sec:CAny}
We are mainly interested in codimension-one observables within the class of the C=Anything proposal, introduced in \cite{Belin:2021bga,Belin:2022xmt},
\begin{equation}\label{eq:Volepsilon}
    \mathcal{C}^\epsilon \equiv \frac{1}{G_N}\int_{\Sigma_\epsilon}\rmd^d\sigma\,\sqrt{h}~F[g_{\mu\nu},\,\mathcal{R}_{\mu\nu\rho\sigma},\,\nabla_\mu]~,
\end{equation}
where $F[g_{\mu\nu},\,\mathcal{R}_{\mu\nu\rho\sigma},\,\nabla_\mu]$ is an arbitrary scalar functional of $d+1$-dimensional bulk curvature invariants, $\Sigma_\epsilon$ is a $d$-dimensional spatial slice labeled by $\epsilon$($=+,\,-$), which is anchored on the stretched horizon, $h$ is the determinant of the induced metric, $h_{\mu\nu}$, on $\Sigma_\epsilon$. We will let $F[\dots]$ be a general functional throughout the work. The reader is referred to footnote \ref{footnote} to verify that these proposals display a switchback effect in AdS planar black holes in the time-reversal symmetrization in (\ref{eq:C min}).

\begin{figure}[t!]
    \centering
    \includegraphics[width=0.324\textwidth]{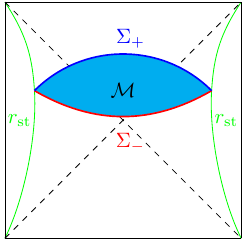}\hspace{0.5cm}\includegraphics[width=0.62\textwidth]{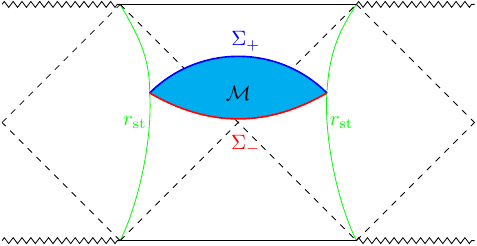}
    \caption{Implementation of the codimension-one CAny proposals with CMC slices, $\Sigma_+$ (blue) and $\Sigma_-$ (red) in the bulk region $\mathcal{M}$ (cyan), as evaluation regions in the unperturbed (S)dS space, following the same notation as in Fig. \ref{fig:SW SdS}.}
    \label{fig:CAny SdS}
\end{figure}

To define the region of evaluation, we employ a combination of codimension-one and codimension-zero volumes with different weights, given by
\begin{equation}\label{eq:regions CMC}
\begin{aligned}
    \mathcal{C}_{\rm CMC}=\frac{1}{G_N}\biggl[&\alpha_+\int_{\Sigma_+}\rmd^d\sigma\,\sqrt{h}+\alpha_-\int_{\Sigma_-}\rmd^d\sigma\,\sqrt{h}+\alpha_B\int_{\mathcal{M}}\rmd^{d+1}x\sqrt{-g}\biggr]
\end{aligned}
\end{equation}
where $\mathcal{M}$ is the bulk region; $\alpha_\pm$, $\alpha_B$ are constants; and $\Sigma_+$, $\Sigma_-$ are the future and past boundary slices in $\partial\mathcal{M}=\Sigma_+\cup\Sigma_-$, see Fig \ref{fig:CAny SdS}. 
The extremization of $\mathcal{C}_{\rm CMC}$ reveals that $\Sigma_\pm$ are CMC slices, whose mean curvature is given by:
\begin{equation}
    K_\epsilon\equiv\eval{K}_{\Sigma_\epsilon} = -\epsilon\frac{\alpha_B}{\alpha_\epsilon}~,
\end{equation}
where $K^{\mu\nu}=h^{\mu\alpha}\nabla_\alpha n^\nu$ is the extrinsic curvature, and we consider $n^\mu$ to be a future pointing normal vector for both $\Sigma_\epsilon$.

To simplify the evaluation of (\ref{eq:Volepsilon}), we employ time-symmetric evolution on each of the static patches, so that we set $t_L=t_R$ in (\ref{metric1}). Moreover, we introduce Eddington-Finkelstein coordinates in (\ref{metric1}),
\begin{equation}\label{eq:starting metric}
    \rmd s^2=-f(r)\rmd v^2+2\rmd v\rmd r+r^2\rmd\Omega^2_{d-1}~,
\end{equation}
which are related to the Kruskal coordinates (\ref{eq:Kruskal}) by
\begin{equation}\label{eq:Kruskal coord}
    U=\rme^{-\frac{f'(r_c)}{2}u}~,\quad V=-\rme^{\frac{f'(r_c)}{2}v}~.
\end{equation}
Evaluating (\ref{eq:Volepsilon}, \ref{eq:regions CMC}) with (\ref{metric1}), one finds
\begin{align}\label{eq:C epsilon inter}
    \mathcal{C}^\epsilon &=\frac{\Omega_{d-1}}{G_N}\int_{\Sigma_\epsilon}\rmd\sigma\,r^{d-1}\sqrt{-f(r)\dot{v}^2+2\dot{v}\dot{r}}\,a(r)\,,\\
    \mathcal{C}_{\rm CMC}&=\frac{\Omega_{d-1}}{G_N}\sum_{\epsilon}\alpha_\epsilon\int_{\Sigma_\epsilon}
    \rmd\sigma\,\mathcal{L}_\epsilon~,\label{eq:C CMC}
\end{align}
where $a(r)$ is a scalar functional corresponding to the evaluation of $F[g_{\mu\nu},\,\mathcal{R}_{\mu\nu\rho\sigma},\,\nabla_\mu]$; $\sigma$ is a general parametrization of the coordinates $v(\sigma)$, $r(\sigma)$ on the slice $\Sigma_\epsilon$; and
\begin{equation}
    \mathcal{L}_\epsilon\equiv r^{d-1}\sqrt{-f(r)\dot{v}^2+2\dot{v}\dot{r}}-\epsilon\frac{K_\epsilon}{d}\dot{v}r^d~.
\end{equation}
The details of the evaluation are shown in App. \ref{App:details}. The late-time evolution of complexity results in:
\begin{equation}\label{eq:Vol late times}
    \lim_{t\rightarrow\infty}\dv{t} \mathcal{C}^\epsilon \simeq \frac{\Omega_{d-1}}{G_N}\sqrt{-f(r_{f}) r_{f}^{2(d-1)}}\,a(r_{f})~\text{ with }r_{f}\equiv\lim_{t\rightarrow \infty}r_{t}~.
\end{equation}
Here $r_f$ is a local maximum of the effective potential at late times:
\begin{equation}\label{eq:extr potential}
    \eval{\mathcal{U}}_{r_{f}}=0,\quad \eval{\partial_{{r}}\mathcal{U}}_{r_{f}}=0,\quad \eval{\partial^2_{{r}}\mathcal{U}}_{r_{f}}\leq0~.
 \end{equation}
These conditions lead to the following relation
\begin{equation}\label{eq:rt location}
   W(r_f,\,K_\epsilon)\equiv 4 r_{f}f\left(r_{f}\right)\left((d-1)f'\left(r_{f}\right)+K_\epsilon^2 r_{f}\right)+4(d-1)^2f\left(r_{f}\right){}^2+r_{f}^2f'\left(r_{f}\right){}^2=0~.
\end{equation}
{The roots $r_f$ of the function $W(r_f,\,K_\epsilon)$ can be found explicitly for pure dS and extremal SdS black hole limits, as originally derived in \cite{Aguilar-Gutierrez:2023zqm},} 
\begin{align}
\label{eq:rf dS}
    \qty({r}^{\rm (dS)}_f)^2 &= \frac{{K_\eps}^2-2 d(d-1)\pm|{K_\eps}| \sqrt{{K_\eps}^2-4(d-1)}}{2( {K_\eps}^2- d^2)}~,\quad\abs{K_\eps}\geq 2\sqrt{d-1}~;\\\label{eq:rf N}
    {r}^{\rm (N)}_f&=\sqrt{\frac{d-2}{d}}~,\quad\abs{K_\eps}\geq\sqrt{d}~.
\end{align}
Let us now show that for a generic SdS black hole spacetimes ($d\geq3$), there will always be a real root to (\ref{eq:rt location}). One can evaluate $W(r_f,\,K_\epsilon)$ in (\ref{eq:rt location}) with the roots in (\ref{eq:rf dS}, \ref{eq:rf N}) while keeping the mass of the black hole arbitrary. We will denote $m\equiv \mu/\mu_N\in[0,\,1]$, such that we may express:
\begin{align}
    &W\qty({r}^{\rm (dS)}_f,\,2\sqrt{d-1})=\frac{4 m  \left((d-2)^d d^2m-4 (d-2)^2 ((d-1) d)^{d/2}\right)}{(d-2)^{4-d} (d-1)^{d-2} d^{d}}~,\label{eq:HdS}\\
    &W\qty({r}^{\rm (N)}_f,\,K_\epsilon)=\frac{4(1-m)\qty(2K_\epsilon^2(d-2)+d^2(1-m))}{d^2}~.\label{eq:HN}
\end{align}
Notice that (\ref{eq:HdS}) is clearly negative for all $d\geq3$ and $m\in(0,\,1)$, while (\ref{eq:HN}) is positive. Moreover, as we increase $\abs{K_\epsilon}>2\sqrt{d-1}$, $W\qty({r}^{\rm (dS)}_f,\,K_\epsilon)$ becomes more negative in (\ref{eq:rt location}). Then, according to the \emph{intermediate value theorem}, there will exist at least a real root ${r}_f\in \qty[{r}^{\rm (dS)}_f,\,{r}^{\rm (N)}_f]$ for general SdS$_{d+1}$ space.

On the other hand, since we have allowed $a(r)$ to be an arbitrary function in (\ref{eq:Vol late times}), we see that when $r_f\rightarrow\infty$\footnote{This condition is satisfied in (\ref{eq:rf dS}) when $K_\epsilon=d$ in dS$_2$ and (S)dS$_3$ space.} there would be arbitrary types of late-time growth for $C^\epsilon$ depending on the particular choice of $a(r)$. For instance, the case $a(r)=1$ leads to late-time exponential behavior when $r_f\rightarrow\infty$; meanwhile, having a different degree of divergence in (\ref{eq:Vol late times}) would lead to enhancement or decrease in the late-time growth. However, for this to be a valid CAny proposal, we require also a modification, as explained below.

When one evaluates the early time evolution in (\ref{eq:Vol late times}) $t\rightarrow -\infty$, there is a sign flip in $K_\epsilon\rightarrow-K_\epsilon$. As a result, the rate of growth of the CAny observables at early and late times does not coincide for a given CMC slice. The future or past growth would be given by (\ref{eq:Vol late times}), while the other generates hyperfast growth.\footnote{See \cite{Aguilar-Gutierrez:2023zqm} for comments about possible interpretations in terms of circuit complexity.} Importantly for us, the switchback effect is not respected in this case, as on requires a cancellation between early and late-time contributions to the complexity growth. We will make this more explicit below.

\section{The switchback effect}\label{sec:switchback}
\begin{figure}[t!]
    \centering
    \begin{minipage}{.5\textwidth}
        \centering
        \includegraphics[height=0.25\textwidth]{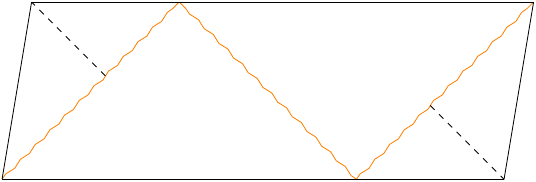}\vspace{0.5cm}\\
        \includegraphics[height=0.25\textwidth]{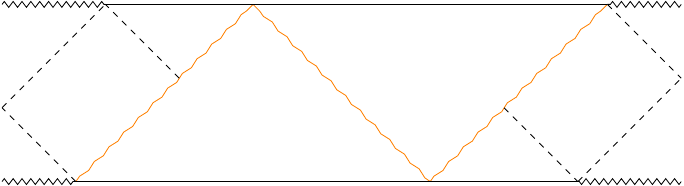}
    \end{minipage}\begin{minipage}{0.5\textwidth}
        \centering
        \includegraphics[height=0.25\textwidth]{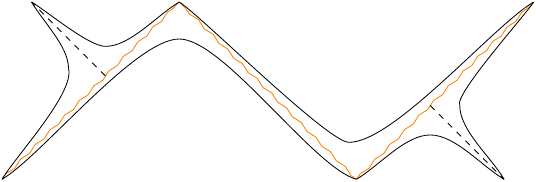}\vspace{0.5cm}\\
        \includegraphics[height=0.25\textwidth]{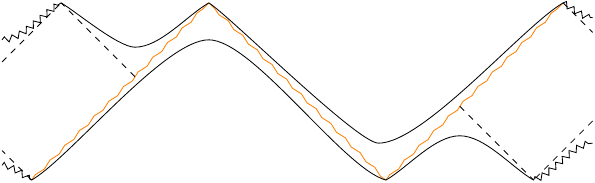}
    \end{minipage}
    \caption{Multiple shockwave geometry in SdS$_{3}$ space (above) SdS$_{d+1\geq4}$ space (below). \textit{Left column}: (\ref{eq:conformal coord}) coordinates, and \textit{right column}: (\ref{eq:Kruskal}) coordinates. The figure shows two forward-evolving and one backward-evolving pulse, producing the corresponding shift (\ref{eq:scrambling cosmo}).}
    \label{fig:many SWs}
\end{figure}

We will study the set of observables (\ref{eq:Volepsilon}, \ref{eq:regions CMC}) in the shockwave geometry (\ref{eq:metric SW alpha}). We begin performing a sequence of an even number of shockwaves, $n$, in the inflating patch (i.e. $r\in[r_{\mathcal{O}},\,\infty]$). Let us denote ${t}_1$, ${t}_2$, \dots, ${t}_n$ as the insertion static patch times with respect to the stretched horizon in alternating insertion order, i.e. ${t}_{2k+1}>{t}_{2k}$, and ${t}_{2k}<{t}_{2k-1}$, restricted to $\abs{{t}_{i+1}-{t}_i}\gg {t}_*$. {Fig. \ref{fig:many SWs} illustrates the multiple shockwave configuration in SdS space. The reader is referred to \cite{Shenker:2013pqa, Stanford:2014jda} for the asymptotic AdS black hole counterpart.}

\begin{figure}
    \centering
    \includegraphics[width=0.32\textwidth]{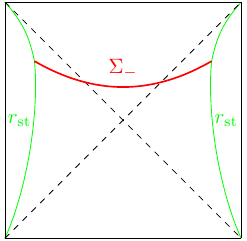}\hfill\includegraphics[width=0.32\textwidth]{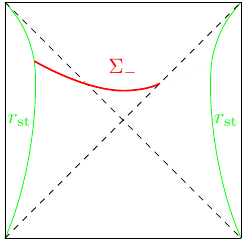}\hfill\includegraphics[width=0.32\textwidth]{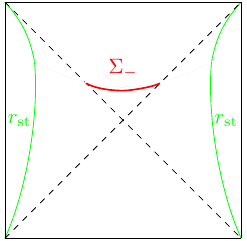}
    \caption{Different configurations of the extremal complexity surface $\Sigma_\epsilon$ appearing in (\ref{eq:multiple SW}) and illustrated for $\epsilon=-$ in SdS$_3$ space. Left: $\mathcal{C}^-(t_L,\,t_R)$, center: $\mathcal{C}^-(V_R,~t_L)$, and right: $\mathcal{C}^-(V_R,~U_L)$.}
    \label{fig:anchoring}
\end{figure}

Accounting for the sign of the shift in the backreacted metric (\ref{eq:metric SW alpha}), the functional (\ref{eq:Volepsilon}) has an additive property under these insertions in the strong shockwave limit \cite{Belin:2021bga,Belin:2022xmt},
\begin{equation}\label{eq:multiple SW}
\begin{aligned}
    {\mathcal{C}}^\epsilon(t_L,\,t_R)=&\mathcal{C}^\epsilon(t_R,\,V_1)+\mathcal{C}^\epsilon(V_1-\alpha_1,\,U_2)+\dots\\
    &+\mathcal{C}^\epsilon(U_{n-1}+\alpha_{n-1},\,V_{n})+\mathcal{C}^\epsilon(V_{n}-\alpha_{n},\,t_L)
\end{aligned}
\end{equation}
where all the contributions $\mathcal{C}^\epsilon(\cdot,\,\cdot)$ follow the same definition (\ref{eq:Volepsilon}), but $\Sigma_\epsilon$ is anchored between endpoints that are located either on the left/right cosmological horizon, $r_c$, or on the stretched horizon $r_{\rm st}$. The different cases are illustrated in Fig. \ref{fig:anchoring}.
We can perform the evaluation of $\mathcal{C}^\epsilon$ by searching the locations $u_{R,\,L}$, $v_{R,\,L}$ where $\Sigma_\epsilon$ intersect with the left/right horizon $r_c$,
\begin{equation}\label{eq:vR}
    v_R-v_t=\int_{r_c}^{r_t}\rmd r\frac{\dot{v}}{\dot{r}}=\int_{r_c}^{r_t}\frac{\rmd r}{f(r)}\qty(1-\frac{P_v^\epsilon+\frac{\epsilon K_\epsilon}{d}r^d}{\sqrt{-\mathcal{U}(P_v\epsilon,\,r)}})~,
\end{equation}
and $v_t=v_R(r_t)$. 

Next, we will use the \textit{boost symmetry} in the static patch to set symmetric time evolution ($t_L=t_R=t/2$, which also implies $u_{L,~R}=v_{R,~L}$). The different contributions in (\ref{eq:multiple SW}) can be expressed with EF coordinates (\ref{eq:starting metric}) as:
\begin{align}
    \mathcal{C}^\epsilon(t_R,\,U_L)=\mathcal{C}^\epsilon(V_R,\,t_L)&=-\frac{\Omega_{d-1}}{G_N}a(r_t)\sqrt{-f(r_t)r_t^{2(d-1)}}\qty(\int_{r_{\rm st}}^{r_t}+\int_{r_c}^{r_t})\frac{\left(P_v^\epsilon + \frac{\epsilon K_\epsilon}{d} r^{d} \right)\rmd r}{f(r)\sqrt{-\mathcal{U}(P_v^\epsilon, r)}}~,\\
    \mathcal{C}^\epsilon(V_R,\,U_L)&=-\frac{2\Omega_{d-1}}{G_N}a(r_t)\sqrt{-f(r_t)r_t^{2(d-1)}}\int_{r_c}^{r_t}\frac{\left(P_v^\epsilon + \frac{\epsilon K_\epsilon}{d} r^{d} \right)\rmd r}{f(r)\sqrt{-\mathcal{U}(P_v^\epsilon, r)}}~.
\end{align}
As outlined in App. \ref{App:details} (see (\ref{eq:Cepsilon close form})), we can express (\ref{eq:C epsilon inter}) as:
\begin{align}\label{eq:C UR VL}
    {\mathcal{C}^\epsilon}(V_R,\,U_L)=&-\frac{2\Omega_{d-1}}{G_N}a(r_t)\sqrt{-f(r_t)r_t^{2(d-1)}}\int_{r_c}^{r_t}\frac{\left(P_v^\epsilon + \frac{\epsilon K_\epsilon}{d} r^{d} \right)\rmd r}{f(r)\sqrt{-\mathcal{U}(P_v^\epsilon, r)}}\\
    &+\frac{2\Omega_{d-1}}{G_N}\int_{r_c}^{r_t}\frac{a(r)f(r)\,r^{2(d-1)}+a(r_t)\sqrt{-f(r_{t})r_{t}^{2(d-1)}}\left(P_v^\epsilon + \frac{\epsilon K_\epsilon}{d} r^{d} \right)}{f(r)\sqrt{-\mathcal{U}(P_v^\epsilon,\,r)}}~.\nonumber
\end{align}
Moreover, one may Taylor expand the effective potential (\ref{eq:extr potential}) around the final slice, $r_f$, as
\begin{equation}\label{eq: Approx Potential}
\lim_{r\rightarrow r_f}\mathcal{U}(P_v^\epsilon,\,r)\simeq \frac{1}{2}(r-r_f)^2\mathcal{U}''(P_v^\epsilon,\,r)+\mathcal{O}(\abs{r-r_f}^3)~.
\end{equation}
We can then evaluate (\ref{eq:vR}, \ref{eq:C UR VL}) with (\ref{eq: Approx Potential}) to find
\begin{equation}\label{eq:Ceps v}
\begin{aligned}
{\mathcal{C}^\epsilon}(V_R,\,U_L)=\frac{\Omega_{d-1}}{G_N}P_\infty^\epsilon v~,\quad P_\infty^\epsilon=a(r_f)\sqrt{-f(r_f)r_f^{2(d-1)}}
\end{aligned}
\end{equation}
where $r_f$ is determined with (\ref{eq:rt location}). The results above (\ref{eq:Ceps v}, \ref{eq:Vol late times}) can be used to express the contributions in (\ref{eq:multiple SW}) in Kruskal coordinates (\ref{eq:Kruskal coord}) as:
\begin{align}
{\mathcal{C}^\epsilon}(t_R,~V_L)&=\frac{\Omega_{d-1}}{G_N}P_{\infty}^\epsilon\log{V_L\rme^{t_R}}~,\\
{\mathcal{C}^\epsilon}(V_R,~U_L)&=\frac{\Omega_{d-1}}{G_N}P_{\infty}^\epsilon\log{U_LV_R}~,\\
{\mathcal{C}^\epsilon}(V_R,~t_L)&=\frac{\Omega_{d-1}}{G_N}P^\epsilon_{\infty}\log{\rme^{t_L}V_R}~.
\end{align}
However, there is also an early time contribution in the shockwave geometry\footnote{This was also noticed in \cite{Aguilar-Gutierrez:2023zqm,silke} for the AdS black hole background.}, given by the term
\begin{equation}
{\mathcal{C}^\epsilon}(V_L,\,U_R)=\frac{\Omega_{d-1}}{G_N}P_{-\infty}^\epsilon\log{V_LU_R}~,
\end{equation}
where
\begin{equation}
    P_{-\infty}^\epsilon=a(r_{I})\sqrt{-f(r_I)r_I^{2(d-1)}}~,~\text{ with }~ r_I=\lim_{t\rightarrow-\infty}r_t~.
\end{equation}
As mentioned above in Sec. \ref{sec:CAny}, for $t\rightarrow -\infty$, there is a sign flip in $K_\epsilon\rightarrow-K_\epsilon$. In that case, $r_I$ is a solution to (\ref{eq:extr potential}, \ref{eq:rt location}) with the appropriate modification of $K_\epsilon$. However, as we also pointed out, the CMC slices that display late time growth in the far past/future display hyperfast growth in the future/past respectively. Instead, consider a protocol where we evaluate (\ref{eq:C epsilon inter}) over different CMC slices in the past and future, such that there are always solutions $r_f$ and $r_I$ with respect to the stretch horizon evolution. (\ref{eq:multiple SW}) then transforms into
\begin{equation}\label{eq:Cepsilon tot}
\begin{aligned}
{\mathcal{C}^\epsilon}(t_L,\,t_R)\simeq \frac{\Omega_{d-1}}{G_N}\Bigl[&P_{\infty}^\epsilon \log(V_1\rme^{t_R})+P_{-\infty}^\epsilon\log(V_1-\alpha_1)U_2+P_{\infty}^\epsilon\log(U_2+\alpha_2)V_3+\dots\\
&+P_\infty^\epsilon\log((V_n-\alpha_n)\rme^{t_L})\Bigr]~.
\end{aligned}
\end{equation}
We can then extremize (\ref{eq:Cepsilon tot}) with respect to an arbitrary interception point ($V_i$, $U_i$) in the multiple shockwave geometry,
\begin{equation}
\dv{{\mathcal{C}^\epsilon}(t_L,\,t_R)}{V_i}=0~,\quad \dv{{\mathcal{C}^\epsilon}(t_L,\,t_R)}{U_i}=0~,
\end{equation}
which allows us to locate
\begin{align}
V_i^\epsilon=\frac{P_{\infty}^\epsilon\alpha_i}{P_\infty^\epsilon+P_{-\infty}^\epsilon}~,\quad U_i^\epsilon=-\frac{P_{-\infty}^\epsilon\alpha_i}{P_\infty^\epsilon+P_{-\infty}^\epsilon}~.
\end{align}
Replacing the interception points into (\ref{eq:Cepsilon tot}) generates:
\begin{equation}\label{eq:Intermediate switchback}
{\mathcal{C}^\epsilon}(t_L,\,t_R)\simeq\frac{\Omega_{d-1}}{G_N}\qty(P_{\infty}^\epsilon (t_R+t_L)+(P_{\infty}^\epsilon+P_{-\infty}^\epsilon)\qty(\sum_{k=1}^nt_k-nt^{(c)}_*))~,
\end{equation}
up to constant terms in terms of $P_{+\infty}^\epsilon$, $P_{-\infty}^\epsilon$.

Importantly, it was noticed in \cite{Aguilar-Gutierrez:2023zqm} that the CAny proposals with a generic functional $F[\dots]$ in (\ref{eq:C CMC}) for an AdS planar black hole background only satisfy the switchback effect when the rate of growth in the past and future are the same. This means that for (\ref{eq:C CMC}) to obey the definition of holographic complexity in \cite{Belin:2021bga,Belin:2022xmt}, we require\footnote{The derivation of this requirement for the AdS planar black hole case follows the same steps that we have presented for the SdS case, although some replacements need to be made. This includes inverting the integration limits in (\ref{eq:vR}, \ref{eq:vR}); setting $r_c\rightarrow r_b$; $K_\epsilon\rightarrow-K_\epsilon$; and $\alpha_i\rightarrow-\alpha_i$.\label{footnote}}
\begin{equation}\label{eq:P epsilon past future}
P^\epsilon_{+\infty}=P^\epsilon_{-\infty}~.
\end{equation}
In that case, the evaluation of (\ref{eq:Intermediate switchback}) reduces to
\begin{equation}\label{eq:switchback result}
    \mathcal{C}^\epsilon(t_L,\,t_R)\propto\abs{t_R+t_1}+\abs{t_2-t_1}+\dots+\abs{t_n-t_L}-2nt^{(c)}_*~,
\end{equation}
where the term $-2n t^{(c)}_*$ appears due to cancellation in the complexity growth due to early and late time perturbations.

\begin{figure}[t!]
    \centering
    \includegraphics[width=0.325\textwidth]{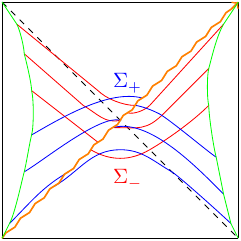}\hspace{0.5cm}
        \includegraphics[width=0.63\textwidth]{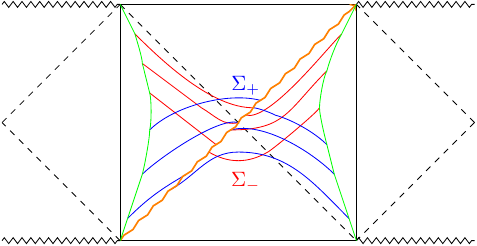}
    \caption{Representative CMC slices anchored to the stretched horizon (green) in a single shockwave geometry in SdS space. We employ the discontinuous Kruskal coordinates $\tilde{U}$, $\tilde{V}$ in (\ref{eq:tilde metric}) in the Penrose diagram to facilitate the representation of the CMC slices.}
    \label{fig:SW CAny SdS}
\end{figure}

Notice that a possible way to satisfy (\ref{eq:P epsilon past future}) in SdS space can be obtained by setting $K_-=-K_+$ and selecting a complexity proposal $\mathcal{C}$ as\footnote{However, instead of minimization, one might as well perform a maximization over the CMC slices, or an averaging, as either of those would satisfy (\ref{eq:P epsilon past future}) in AdS planar black holes; although that would reproduce the hyperfast growth in SdS space.}
\begin{equation}\label{eq:C min}
    \mathcal{C}=\min_t\qty(\mathcal{C}^+(t),\,\mathcal{C}^-(t))~.
\end{equation}
See Fig. \ref{fig:SW CAny SdS} for an illustration of the evolution of the CMC slices in the shockwave geometry.

We close the section with a few remarks. First, the result (\ref{eq:switchback result}) reproduces the same type of behavior as the switchback effect for AdS planar black holes, at least for the CAny proposals with early and late-time linear growth in SdS space. Second, as we mentioned in Sec. \ref{sec:CAny} there are fine-tuned situations where $\mathcal{C}^\epsilon$ can have any type of early and late-time growth behavior for SdS$_{3}$. It might be interesting to study the modifications in the switchback in those cases. Lastly, the switchback effect has also been recovered in a different and more explicit analysis for particular asymptotically dS backgrounds \cite{Anegawa:2023dad,Baiguera:2023tpt}, hinting at the possibility that this is a rather universal phenomenon in shockwave geometries.

\section{Discussion and outlook}\label{sec:conclusions}
In summary, we studied the appearance of the switchback effect in asymptotically dS spacetimes by studying the late (and early time) evolution of the codimension-one CAny observables under shockwave insertions. We picked a set of observables that are evaluated in CMC slices of different curvature in the past and future boundaries. We proved that under a weakly gravitating regime, the CAny observables show a reduction in the complexity growth due to cancellations of the energy perturbations. We also explicitly verified one of the predictions in \cite{Aguilar-Gutierrez:2023zqm}, namely that a time-reversal symmetric protocol would be necessary for the switchback effect to occur\footnote{On the field theory side, the notion of Nielsen geometric approach to complexity \cite{Nielsen,nielsen2006quantum,dowling2006geometry} suggests that this must be indeed respected.}. Moreover, our findings show a great similarity with respect to the behavior of CAny proposals for AdS black holes under the switchback effect. However, we reiterate that the CAny observables in our study do not necessarily represent holographic complexity in asymptotically dS space. To have a clear notion of holographic complexity, we might require a quantum circuit interpretation for the observables, which would also require a reliable quantum circuit model for dS space. Some toy models allowing much progress in this direction have been studied in \cite{Bao:2017iye,Bao:2017qmt,Niermann:2021wco,Cao:2023gkw}.

We comment on some interesting future directions. Our work focused on the alternating shockwave insertion on the inflating patch of the SdS spacetime. However, we can also extend the analysis when the SdS spacetime has multiple patches, to enquire about the black hole interior as well. These types of geometries have been used as toy models for multiverses in \cite{Aguilar-Gutierrez:2021bns}. One of the striking features previously found was that the information available to the past light cone of an observer in one of the multiverses would encode the information of the other universes, in the semiclassical regime. It would be very interesting to see whether the notions of general codimension-zero CAny observables may also encode such information. We might be able to learn if some of these observables might have an interpretation from the point of view of quantum cosmology. It might be also fruitful to study how the introduction of perturbations in the geometry affects the coarse-graining of information found in \cite{Aguilar-Gutierrez:2021bns}.

An important aspect in the search for the holographic description of dS space would be obtaining a dual interpretation of the observables that we studied in this work. It would be interesting to analyze quantum circuit observables that display similar behaviors to the ones studied within our work to study whether the effect of perturbations on the stretched horizon might also have the interpretation of an epidemic type of growth given the insertion of operators. Moreover, it would be interesting to see what type of signals can be found in a UV complete description of the stretched horizon, motivated by the DSSYK model \cite{Susskind:2021esx,Blommaert:2023opb}.

Applications to of CAny to dS space braneworld models were recently carried out in \cite{Aguilar-Gutierrez:2023tic} to gain more information about dS holography. In these models \cite{Randall:1999ee,Randall:1999vf}, one includes an end-of-the-world brane, whose tension determines the cosmological constant in the effective gravitational theory on the brane. It would be interesting to incorporate the switchback effect to characterize perturbations in a double holographic setting, with a clearer field theory dual. This effective theory might be further modified by adding intrinsic gravity theories on the brane, leading to a more intricate holographic complexity evolution \cite{Hernandez:2020nem,ACHKKS,AL}. Moreover, the fluctuations associated with the brane location lead to an effective description as dS JT gravity on one of the branes \cite{Aguilar-Gutierrez:2023tic}. It would be interesting to study the switchback effect of this effective theory.

On the other hand, as we found in Fig. \ref{fig:CAny SdS}, the CAny observables generically reach a terminal turning $r_f$ (as well as a time-reversal version) determined by the choice of the CMC slices through (\ref{eq:rt location}). This implies that the CAny proposals in the article do not prove the whole cosmological patch of SdS spacetime. However, we would expect that any notion of static patch holography should also encode the degrees of freedom of the inflating region, similar to investigations in asymptotically AdS space \cite{Jorstad:2023kmq}.\footnote{We thank Eivind Jorstad for correspondence about this point.} Nevertheless, one can probe more of the geometry outside the cosmological horizon using the alternating shockwave geometry. It seems that adding perturbations in the stretched horizon reveals more information, even when its explicit localization is irrelevant.

Finally, much progress in the dS holography has been made possible through the study of interpolating geometries in two-dimensional dilaton-gravity theories \cite{Anninos:2017hhn,Anninos:2018svg,Anninos:2020cwo,Anninos:2022hqo,Aguilar-Gutierrez:2023odp,AESV}. While the CV proposal has been extensively studied in \cite{Chapman:2021eyy} for certain interpolating geometries; new members in this set have recently appeared \cite{Anninos:2022hqo}, and the CAny proposals have not been treated yet. This might allow for a clearer interpretation of the properties studied in this work in the context of dS$_2$ space, appearing from near extremal limits near horizon limits of black hole geometries
\cite{Maldacena:2019cbz,Castro:2022cuo}.

\section*{Acknowledgements}
It's my pleasure to thank Stefano Baiguera, Alex Belin, Rotem Berman, Michal P. Heller, Eivind Jørstad, Edward K. Morvan-Benhaim, Juan F. Pedraza, Andrew Svesko, Silke Van der Schueren, and Nicolò Zenoni for valuable discussions. I also thank the University of Amsterdam, and the Delta Institute for Theoretical Physics for their hospitality and support during various phases of this project; and to the organizers of the Modave summer school, where part of the project was completed. The work of SEAG is partially supported by the FWO Research Project G0H9318N and the inter-university project iBOF/21/084.

\appendix

\section{Details on the CAny evaluation}\label{App:details}
For the choice
\begin{equation}
\sqrt{-f(r)\dot{v}^2+2\dot{v}\dot{r}}=r^{d-1}~,
\end{equation}
the Euler-Lagrange equations corresponding to (\ref{eq:C CMC}) can be expressed as
\begin{equation}
    \dot{r}^2+\mathcal{U}(P_v^\epsilon,\,r)=0~,
\end{equation}
where
\begin{align}\label{eq:potential arbitrary}
P_v^{\epsilon}&\equiv\pdv{\mathcal{L}_\epsilon}{\dot{v}}=\dot{r}-\dot{v}\,f(r)-\epsilon\frac{K_\epsilon}{d}\,r^d~;\\
    \mathcal{U}(P_v^{\epsilon},\,r)&\equiv -f(r) r^{2(d-1)} - \left(P_v^\epsilon +\epsilon \frac{K_\epsilon}{d}  r^d\right)^2~.\label{eq:U pot}
\end{align}
Then, we can express (\ref{eq:C epsilon inter}) as
\begin{equation}\label{eq:Cepsilon car}
    \mathcal{C}^\epsilon=\frac{2\Omega_{d-1}}{G_N}\int_{r_{\rm st}}^{r_{t}}\frac{r^{2(d-1)}a(r)}{\sqrt{-\mathcal{U}(P_v^\epsilon,\,r)}}\rmd r~.
\end{equation}
In a similar way, the parameter $t$ can be expressed as
\begin{equation}\label{eq:time def}
\begin{aligned}
t&=\int_{\Sigma_\epsilon} \rmd r\frac{\dot{t}}{\dot{r}}=\int_{\Sigma_\epsilon} \rmd r\frac{\dot{v}-\dot{r}/{f(r)}}{\sqrt{-\mathcal{U}(P_v^\epsilon,\,r)}}\\
    &= -2\int_{r_{\rm st}}^{r_{t}} \frac{\rmd r}{f(r)\,\sqrt{-\mathcal{U}(P_v^\epsilon, r)}} \left(P_v^\epsilon + \frac{\epsilon K_\epsilon}{d} r^{d} \right) ~.
\end{aligned}
\end{equation}
We proceed to evaluate (\ref{eq:Cepsilon car}) with (\ref{eq:time def}) carefully. Since $U(P_v^\epsilon,\,r_{t})=0$ by definition, we need to take care of the denominator in (\ref{eq:Cepsilon car}, \ref{eq:time def}) at each of the turning points. We do so by adding a subtracting a term:
\begin{align}\label{eq:Cepsilon close form}
    {\mathcal{C}^\epsilon}=&-\frac{2\Omega_{d-1}}{G_N}a(r_t)\sqrt{-f(r_t)r_t^{2(d-1)}}\int_{r_{\rm st}}^{r_{t}}\frac{\left(P_v^\epsilon + \frac{\epsilon K_\epsilon}{d} r^{d} \right)\rmd r}{f(r)\sqrt{-\mathcal{U}(P_v^\epsilon, r)}}\\
    &+\frac{2\Omega_{d-1}}{G_N}\int_{r_{\rm st}}^{r_{t}}\frac{a(r)f(r)\,r^{2(d-1)}+a(r_t)\sqrt{-f(r_{t})r_{t}^{2(d-1)}}\left(P_v^\epsilon + \frac{\epsilon K_\epsilon}{d} r^{d} \right)}{f(r)\sqrt{-\mathcal{U}(P_v^\epsilon,\,r)}}~.\nonumber
\end{align}
Then, we can identify the relationship between time in (\ref{eq:time def}) and complexity in (\ref{eq:Cepsilon close form})
\begin{equation}
\begin{aligned}
    {\mathcal{C}^\epsilon}=&\frac{\Omega_{d-1}}{G_N}a(r_t)\sqrt{-f(r_{t})r_{t}^{2(d-1)}}t\\
    &+\frac{2\Omega_{d-1}}{G_N}\int_{r_{\rm st}}^{r_{t}}\frac{a(r)f(r)\,r^{2(d-1)}+a(r_t)\sqrt{-f(r_{t})r_{t}^{2(d-1)}}\left(P_v^\epsilon + \frac{\epsilon K_\epsilon}{d} r^{d} \right)}{f(r)\sqrt{-\mathcal{U}(P_v^\epsilon,\,r)}}~.
\end{aligned}
\end{equation}
We can then straightforwardly take the time derivative,
\begin{align}
   {\dv{\mathcal{C}^\epsilon}{t}}=&\frac{\Omega_{d-1}}{G_N}a(r_t)\sqrt{-f(r_{t})r_{t}^{2(d-1)}}\label{eq:complexity growth}\\
    &+\frac{2\Omega_{d-1}}{G_N}\dv{P_v^\epsilon}{t}\int_{r_{\rm st}}^{r_{t}}\rmd r\frac{
        r^{2(d-1)} \left(a(r_t)\sqrt{-f(r_{t}) r_{t}^{2(d-1)}} 
        - a(r)\qty(P_v^\epsilon + \frac{\epsilon K_\epsilon}{d} r^{d})\right)}{\qty(-\mathcal{U}(P_v^\epsilon,\,r))^{3/2}}~.\nonumber
\end{align}
Thus, in the $t\rightarrow\infty$ regime we then recover (\ref{eq:Vol late times}). Moreover, the $\dv{P_v^\epsilon}{t}$ term vanishes (\ref{eq:Vol late times}) provided that the effective potential reaches a maximum \cite{Jorstad:2023kmq}, shown in (\ref{eq:extr potential}).

\bibliographystyle{JHEP}
\bibliography{references.bib}
\end{document}